\documentclass[final,times,twocolumn]{elsarticle}
\usepackage{lineno,hyperref}
\modulolinenumbers[5]
\usepackage{natbib}
\usepackage{tabularx}
\usepackage{graphicx}
\usepackage{amsmath,stackrel,amssymb}
\usepackage{xspace}
\usepackage{comment}
\journal{Parallel Computing}

\usepackage{listings}
\usepackage{xcolor}

\definecolor{codegreen}{rgb}{0,0.6,0}
\definecolor{codegray}{rgb}{0.5,0.5,0.5}
\definecolor{codepurple}{rgb}{0.58,0,0.82}
\definecolor{backcolour}{rgb}{0.95,0.95,0.92}

\lstdefinestyle{mystyle}{
    commentstyle=\color{codegreen},
    keywordstyle=\color{magenta},
    numberstyle=\tiny\color{codegray},
    stringstyle=\color{codepurple},
    basicstyle=\ttfamily\scriptsize,
    breakatwhitespace=false,         
    breaklines=true,                 
    captionpos=b,                    
    keepspaces=true,                 
    showspaces=false,                
    showstringspaces=false,
    showtabs=false,                  
    tabsize=2
}

\newcommand{\HPCToolkit}{\texttt{HPCToolkit}\xspace}

\newcommand{\nvprof}{\texttt{nvprof}\xspace}
\newcommand{\HiOp}{{HiOp}\xspace}

\newcommand{\remove}[1]{}

\begin{document}

\begin{frontmatter}
\title{
Porting the Nonlinear Optimization Library HiOp to Accelerator-Based Hardware Architectures
}
\author[pnnl]{Slaven Pele\v{s}}
\author[ornl]{Kalyan S. Perumalla}
\author[ornl]{Maksudul Alam}
\author[pnnl]{Asher J. Mancinelli}
\author[pnnl]{R. Cameron Rutherford}
\author[pnnl]{Jake Ryan}
\author[llnl]{Cosmin G. Petra}

\address[pnnl]{Pacific Northwest National Laboratory}
\address[ornl]{Oak Ridge National Laboratory}
\address[llnl]{Lawrence Livermore National Laboratory}

\date{\today}
\begin{abstract}
While interior point methods have been the centerpiece of nonlinear programming tools used in science and engineering, their reliance on linear solvers that can tackle sparse symmetric indefinite and highly ill-conditioned problems made it difficult to implement them effectively on hardware accelerators. At this time, there are few sparse linear solvers that can be used in this context \cite{swirydowicz2022linear}.
Here, we present a novel formulation of an interior point method implemented in our HiOp library, which is designed to be able to run entirely on hardware accelerators. This formulation avoids dependence on sparse solvers altogether, which is achieved by compressing the underlying sparse linear problem into a dense one of manageable size. We demonstrate feasibility of this approach and provide a baseline for future interior point method implementations on hardware accelerators. Our investigation is motivated by problems arising in optimal power flow analysis in power systems engineering and our approach is tailored to the broad class of problems arising in that important domain. We also demonstrate utility of modern programming models based on performance portability libraries, namely, Umpire and RAJA. We discuss trade-offs between performance, portability and development cost in the solution space for this non-linear optimization problem. As a result of this research, we demonstrate for the first time that interior point methods for sparse problems can be efficiently realized on modern computing systems where more than 90\% of processing power is in GPUs.
\end{abstract}

\begin{keyword}
Solvers, optimization, nonlinear programming, GPU, power grids, hardware abstraction, portability.
\end{keyword}
\end{frontmatter}

\section{Introduction}

Interior point methods have been widely used in solving nonlinear programming problems arising in a number of science and engineering areas including model predictive control \cite{biegler2013survey},
structural mechanics \cite{petra2019memory}, genomics \cite{duan2010three}, and optimal power flow
analysis \cite{kai1999interior}. The computational performance of an actual interior point method implementation
strongly depends on the linear solver used to solve the underlying Karush-Kuhn-Tucker (KKT)
linear problem \cite{wachter2006implementation}. Those problems are symmetric and typically indefinite, sparse, and
very ill-conditioned. Many interior point method implementations also require a
linear solver to provide matrix inertia (i.e.~the number of positive, zero, and negative
eigenvalues). 

\begin{figure}
    \centering
    \includegraphics[width=0.8\columnwidth]{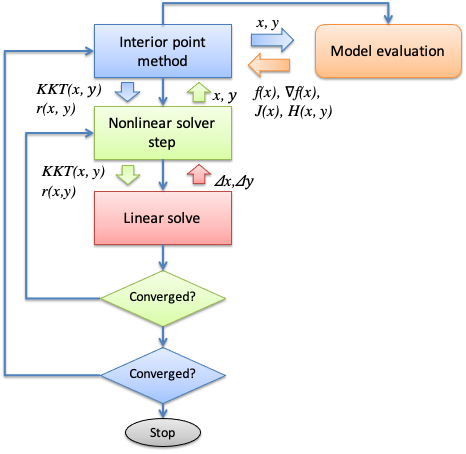}
    \caption{Interior point method workflow. Line arrows describe optimization algorithm control instructions, while block arrows describe data movement between solver components and the optimization model. The method requests evaluation of model objects, assembles KKT system and drives execution of the Newton solver. }
    \label{fig:workflow}
\end{figure}

The interior point method workflow we consider in this paper is schematically shown in Figure~\ref{fig:workflow}. The
method iteratively searches for an optimal solution by minimizing the objective scalar function that depends
on multiple parameters. To ensure that the solution is within a feasible region, that is, all constraints are satisfied, the method extends the problem with barrier functions. Such an extended problem is solved
by a Newton solver (inner loop in Figure~\ref{fig:workflow}), and then barrier parameters are
gradually reduced to small values by a homotopy (outer loop in Figure~\ref{fig:workflow})
to arrive at the solution of the original problem.
Because the underlying linear problem becomes singular when barrier parameters are zero, the
homotopy algorithm has to exit when the optimal solution is close enough to the solution
of the original problem, but before the underlying linear problem becomes too ill-conditioned to solve.
This is why a robust linear solver is essential for an effective implementation of the interior point method. For more details, the reader is referred to seminal paper  \cite{wachter2006implementation} and references therein.

\subsection{Motivating Application}

In the energy industry, optimal power flow analysis is used to minimize the cost of operation
of power grid, while satisfying all operational constraints such as generation-load balance, transmission
limits, frequency limits, etc. The state-of-the-art in the industry is to linearize the power grid
models around the operating point and then use linear programming to get an optimal solution.
This approximation excludes important details such as modeling of voltage and reactive power
constraints. Interior point method captures these constraints and could potentially allow for
less conservative and thus cheaper dispatch within the same safety margin. 

The motivating problem for this work is {\em security constrained} optimal power flow.
In this analysis, one performs simultaneous optimal power flow analysis for multiple scenarios,
where each scenario represents a failure of a selected grid component. Such an analysis provides a contingency
plan for grid planners and operators \cite{chakrabarti2014security,petra2014real}.
In this context, the interior point solver forms a coarse computational unit of a large-scale analysis.
Such analysis is computationally extremely expensive; grid operators, for example, typically
first simplify the problem by reducing the number of contingencies using some heuristic screening
technique, and then perform security constrained {\em power flow} analysis. That means they look for a secure solution
instead of a secure {\em optimal} solution.  

The recent emergence of heterogeneous computing platforms offers new potential to provide the necessary computational power
to enable the energy industry to perform high-fidelity security constrained optimal power flow analysis
effectively on commodity hardware, thereby progressing from merely feasible solutions towards optimal solutions. Deploying the interior point method on an accelerator device such as a graphical processing unit (GPU)
would make a critical step in that direction. 

Historically, interior point methods came to prominence {\em after} efficient sparse linear
solver methods were invented. Challenges in developing efficient fine-grained parallelization techniques for direct sparse linear solvers are currently preventing efficient implementations of interior point method on hardware accelerators.
This limitation is even
more pronounced in the case of highly irregular and very sparse problems, such as power grids, where linear solvers
employing supernodal matrix factorization are not quite effective~\cite{booth2016basker}.
Currently, there are few linear solvers that meet the requirements of the interior point method and
perform well on GPUs \cite{swirydowicz2022linear}.

\subsection{Our Approach}

HiOp is our lightweight optimization solver specifically designed to overcome this issue by compressing
the optimization problem in way that the underlying linear problem is dense and, thus, one can use dense GPU-enabled linear solvers which are more mature and better performing on accelerator devices. This compression technique reduces the size of the problem by approximately a third and is elaborated in Section~\ref{sec:mds}. 

We also use the Kron reduction technique~\cite{DorflerBullo_13_KronRed} to perform an initial, pre-optimization compression of the sparse power flow to dense power flow and to obtain a so-called mixed dense-sparse format used by HiOp to further compress the internal linear systems, as previously mentioned. Kron reduction is a traditional technique in circuit analysis that reduces the number of nodes in a network while maintaining an accurate representation of the physics of the power flow~\cite{DorflerBullo_13_KronRed}. Kron reduction is  essentially equivalent to taking the Schur complement of a subset of the nodes in the susceptance matrix, which, as expected, results in a much denser (and smaller) susceptance matrix. We use the dense Kron-compressed power flow as the power flow engine in the ACOPF formulation and gradually enforce thermal limits on transmission elements and voltage limits by adding additional constraints on the reduced network to the Kron ACOPF problem~\cite{DorflerBullo_13_KronRed}. The Kron compression typically reduces the optimization problem by half for realistic power grid models. We observed that the compression depends on grid connectivity, i.e.~how generators and loads are connected to buses.

With the combined two-stage compression we can use a dense linear solver in optimal power flow analysis for grids with 4000-5000 buses, what is more than twice the size of Texas grid, and fit that computation on a single NVIDIA V100 GPU. For larger grid models currently available ({\em e.g.},~20,000-bus New England grid) we would need a better compression technique and/or to distribute computation over several accelerator devices. This is, however, beyond the scope of this paper as we here focus primarily on feasibility assessment of the mixed dense-sparse approach.

The rest of the paper is organized as follows: 
In Sec.~\ref{sec:mds} we describe novel mixed dense-sparse interior point method formulation. In Sec.~\ref{sec:porting} we provide details on the implementation, while in Sec.~\ref{sec:perfeval} we provide preliminary performance results. We summarize our findings in Sec.~\ref{sec:conclusion}.

\section{Mixed Dense-Sparse Linear Algebra for Interior-Point Methods}~\label{sec:mds}

In this work we consider nonlinear, possibly non-convex optimization problems that explicitly partitions the optimization variables into so-called ``dense'' and ``sparse'' variables, $x_d$ and $x_s$, respectively. 
The optimization problem can be expressed compactly as
\begin{align}
\min_{x_d\in\mathbb{R}^{n_d}, x_s\in\mathbb{R}^{n_s}} & \hspace{0.3cm} f(x_d, x_s) && \label{obj}\\
\textnormal{s.t.} &\hspace{0.3cm}  g(x_d, x_s)=g_E, \label{coneq}&&\\
 &\hspace{0.3cm} h^l \leq h(x_d, x_s) \leq h^u,  &&\label{ineq} \\
 &\hspace{0.3cm} x^{l}_{d} \leq x_d \leq x^{u}_{d}, \ x^{l}_{s} \leq x_s \leq x^{u}_{s}. &&\label{bounds}
\end{align}
Here $f:\mathbb{R}^n\rightarrow\mathbb{R}$, $g:\mathbb{R}^n\rightarrow\mathbb{R}^{m_E}$, and $h:\mathbb{R}^n\rightarrow\mathbb{R}^{m_I}$, where $n$ denotes the total number of variables, $n=n_d + n_s$. The bounds appearing in the inequality constraints~\eqref{ineq} are assumed to be $h^l\in\mathbb{R}^{m_I}\cup\{-\infty\}$, $h^u\in\mathbb{R}^{m_I}\cup\{+\infty\}$, $h_i^l < h_i^u$, and at least of one of $h_i^l$ and $h_i^u$ are finite for each $i\in\{1,\ldots,m_I\}$. The vector bounds $x^{l}_{d}$, $x^{u}_{d}$, $x^{l}_{s}$, and $x^{u}_{s}$ in~\eqref{bounds} need to satisfy identical requirements.  
For the rest of the paper $m$ will denote $m_E+m_I$, \textit{i.e.}, the total number of constraints excepting the simple bounds constraints~\eqref{bounds}.

In addition, we make two assumptions on the block structure of the derivatives that hold for optimal power flow analysis -- our target application -- but not necessarily for an arbitrary nonlinear optimization problem:
\begin{itemize}
    \item[$A_1$.] The ``cross-term'' Hessian matrices $\nabla^2_{x_d x_s} f$, $\nabla^2_{x_s x_d} f$, $\nabla^2_{x_d x_s} g$, $\nabla^2_{x_s x_d} g$, $\nabla^2_{x_d x_s} h$, and $\nabla^2_{x_s x_d} h$ are zero;
    \item[$A_2$.] The Hessian matrix $\nabla^2_{x_s x_s} L$ has a sparsity pattern that allows \textit{computationally efficient} inversion of (or solving with) the matrix $\nabla^2_{x_s x_s} L + D_{x_s}$ where $D_{x_s}$ is a diagonal matrix with positive diagonal entries. In our target applications, namely, optimal power flow problems, $\nabla^2_{x_s x_s} L$ is a diagonal matrix with nonnegative entries.
\end{itemize}

Here, $L$ refers to a Lagrangian function associated with the optimization problem $\eqref{obj}-\eqref{bounds}$, see~\cite{petra2019memory} for its expression, and its Hessian is 
\begin{align*} 
\nabla^2 L(x_d,x_s)  = & \nabla^2 f(x_d,x_s) + \\
&\sum_{i=1}^{m_E} y_{g,i} \nabla^2 g_i(x_d,x_s) + \\ &\sum_{i=1}^{m_I} y_{h,i} \nabla^2 h_i(x_d,x_s),
\end{align*}
where the  vectors $y_g$ and $y_h$ are the Lagrange multipliers corresponding to constraints~\eqref{coneq} and~\eqref{ineq}, respectively.

The salient idea behind mixed dense-sparse problems of the form~\eqref{obj}-\eqref{bounds} is that the explicit partitioning of the optimization variables and the structural properties of the functions $f(\cdot)$, $g(\cdot)$, and $h(\cdot)$, given by assumptions $A_1$ and $A_2$, allow orchestrating the computations of the optimization algorithm to heavily rely on matrix and vector \textit{dense} kernels and to reduce the reliance on sparse linear algebra kernels.

\subsection{The interior-point method loop}\label{sec:ipmalgorithm}

 For this work, HiOp implements the filter line-search interior-point method (IPM) of W{\"a}chter and Biegler~\cite{waecther_05_ipopt,waecther_05_ipopt2} (also implemented by IPOPT~\cite{wachter2006implementation}). 
Paramount to guaranteeing global convergence from remote starting points under mild assumptions~\cite{waecther_05_ipopt} is the use of a line-search procedure equipped with a filter~\cite{Fletcher}.  The basic idea behind this approach is to ($i$) accept trial points that improve the objective function or improve the constraint violation (instead of a combination of
those two measures that is typical for merit- or penalty-based line-searches) and ($ii$) reject trial points that could potentially cause the algorithm to cycle. The line-search backtracks along the search direction until a trial point is accepted or rejected and requires only vector-vector ``axpy'' operations. The search is started at the point that is feasible with respect to the bounds constraints and is the farthest away from the incumbent iterate along the search direction. This operation requires a ``reduce'' operation and has negligible cost relative to the total computational cost.

The optimization problem~\eqref{obj}--\eqref{bounds} is transformed internally by the optimization solver HiOp to an equivalent form that is more amenable to the use of interior-point methods as described in~\cite[Section~3]{petra2019memory}. The search direction, which can be interpreted as the Newton direction for solving the optimality conditions of the log-barrier subproblem employed by the method, is computed by solving a specific class of linear systems~\cite{petra2019memory,wachter2006implementation} in the the form of

{\footnotesize
\begin{align} \label{linsys4x4}
    \left[\begin{array}{cccc}
        Q_{x_s}   & 0         & (J_{x_s} g)^T & (J_{x_s} h)^T \\
        0         & Q_{x_d}   & (J_{x_d} g)^T & (J_{x_d} h)^T \\
        J_{x_s} g & J_{x_d} g & 0           & 0 \\
        J_{x_s} h & J_{x_d} h & 0           & -D_h^{-1} \\
    \end{array}\right]
    \left[\begin{array}{c} \Delta x_s \\ \Delta x_d \\ \Delta y_g \\ \Delta y_h\end{array}\right] = 
    \left[\begin{array}{c}  r_{x_s} \\ r_{x_d} \\ r_{y_g} \\  r_{y_h}\end{array}\right],
\end{align}
}
where $J_{x_s} \equiv \nabla_{x_s}$, $J_{x_d} \equiv \nabla_{x_d}$, $Q_{x_s} = \nabla^2_{x_s x_s} L + D_{x_s}$ and $Q_{x_d} = \nabla^2_{x_d x_d} L + D_{x_d}$. Here, $D_{x_s}$, $D_{x_d}$, and $D_h$ matrices are diagonal with positive diagonal entries, which arise from the use of log-barrier penalties by the interior-point method~\cite{petra2019memory}. For our target application we assume $Q_{x_s}$ is diagonal, as well.

A key requirement for the convergence of the filter line-search algorithm consists of a certain descent property of the search direction $\left[\Delta x_s, \Delta x_d, \Delta y_g, \Delta y_h\right]$~\cite{waecther_05_ipopt}. This property can be mathematically characterized by the condition that the Hessian (the top left $2\times 2$ diagonal block in~\eqref{linsys4x4}) is uniformly positive definite on the null space of the Jacobian of the constraints~\cite{waecther_05_ipopt}. In computational practice, this positiveness condition is checked based on the inertia of the system matrix from~\eqref{linsys4x4}, namely the inertia has to be $(n,0,m)$~\cite{wachter2006implementation}. Moreover, the condition is enforced by modifying the eigenvalues of the matrix  from~\eqref{linsys4x4} via additions of diagonal matrices to the diagonal blocks in~\eqref{linsys4x4}~\cite{wachter2006implementation}. More specifically, this regularization technique adds positive multiples of the identity matrix to the Hessian diagonal blocks $Q_{x_s}$ and $Q_{x_d}$ and negative multiples of the identity to the diagonal blocks $(3,3)$ and $(4,4)$. The regularization is performed repeatedly for increasingly large multiples until the linear solver reports that inertia for~\eqref{linsys4x4} is  $(n,0,m)$.  Section~\ref{sec:complinalg} elaborates how this regularization mechanism can be performed under the compression/reduction technique that we propose in the following subsection.

\subsection{The mixed dense-sparse linear algebra}\label{sec:mdslinalg}


To further compress the linear system~\eqref{linsys4x4}, we use a Gauss elimination of the sparse blocks. One can easily verify that the linear system~\eqref{linsys4x4} is equivalent to solving a linear system involving the matrix

{\footnotesize
\begin{align} \label{eq:compressed}
  \left[\begin{array}{ccc}
    Q_{x_d}   &  (J_{x_d} g)^T                                    &  (J_{x_d} h)^T   \\
    J_{x_d} g & -J_{x_s} g \cdot Q_{x_s}^{-1} \cdot (J_{x_s} g)^T & -J_{x_s} g \cdot Q_{x_s}^{-1} \cdot (J_{x_s} h)^T \\
    J_{x_d} h & -J_{x_s} h \cdot Q_{x_s}^{-1} \cdot (J_{x_s} g)^T & -J_{x_s} h \cdot Q_{x_s}^{-1} \cdot (J_{x_s} h)^T - D_h^{-1}
  \end{array}\right].
\end{align}
}

The reduction offered by the above system is $n_s$ out of $n_s+n_d+m$ and can be considerable (typically around 30\% reduction) for Kron-reduced ACOPF since both $n_d$ and $m$ are comparable to $n_s$. The system above is not only smaller but also denser. Blocks $(1,1)$, $(1,2)$, and $(1,3)$ (and $(2,1)$ and  $(3,1)$ by symmetry) are fully dense, while blocks $(2,2)$, $(2,3)$, and $(3,3)$ are subject to considerable fill-in given by the matrix multiplications required to compute them. 

\subsection{Computational considerations}\label{sec:complinalg}

The computations performed by the optimization solver can be roughly categorized in the following four kernels types:
\begin{itemize}
\item[K1] vector-vector operations involving only dense vectors, for example, in updating the optimization variables along the search direction, computing residual norms, searching for min/max elements, etc.;
\item[K2] matrix-vector operations involving both sparse and dense matrices and dense vectors, for example in residual computation;
\item[K3] additions and multiplications of matrices in the form of $M := M+ADB^T$, where $A$ and $B$ are sparse, $D$ is a diagonal matrix, and $M$ is dense;
\item[K4] solving the compressed dense linear systems of $3\times 3$ block structure as described in (\ref{eq:compressed}).
\end{itemize}
We remark that K3 is needed to perform the compression from the  $4\times 4$ linear system~\eqref{linsys4x4} to the compressed $3\times 3$ linear system (\ref{eq:compressed}), which is  solved as a dense linear system using MAGMA (see K4). 

Since MAGMA uses the Bunch-Kaufman diagonal pivoting method to form a $LDL^T$ factorization~\cite{bunchkaufman,DuffReid}, the inertia of the $3\times 3$ linear system~\eqref{eq:compressed} can be effectively computed from the $D$ matrix (which is block diagonal with blocks of size $1 \times 1$ and $2\times 2$ pivots). Consequently, the Haynsworth inertia additivity~\cite{HAYNSWORTH196873} formula can be used to compute the inertia of~\eqref{linsys4x4}, which is needed for the inertia correction and regularization computations described in Section~\ref{sec:ipmalgorithm}. More specifically, the inertia of the $4 \times 4$ block matrix is given by the sum of the inertia of $Q_{x_s}$ plus the inertia of the $3 \times 3$ matrix from~\eqref{eq:compressed}. 

Model objects, such as Jacobian and Hessian matrices are computed and supplied to HiOp by the external modeling framework. Design of those specialized kernels is described in \cite{abhyankar2021exago}.


\section{Porting HiOp for Execution on GPUs}\label{sec:porting}

Using problem compression to create a dense underlying linear problem of manageable size enabled us to use more mature dense linear solvers from Magma library \cite{tdb10}. This addressed the main computational cost item K4. To get the full benefit of hardware accelerators, we needed to port K1, K2, and K3 kernels for execution on GPUs, as well.

Similar to most iterative methods, the algorithm control logic for interior point method as implemented in HiOp is inherently sequential. 
As shown in the workflow in Figure \ref{fig:workflow}, the next iteration is decided based on the results
from the prior one. The parallel implementation of such code is obtained by parallelizing the linear algebra
primitives that are utilized by the algorithm. This is a common approach for iterative methods and is employed
in other libraries such as Trilinos \cite{trilinos-paper}, SUNDIALS \cite{hindmarsh2005sundials},
and PETSc \cite{petsc-user-ref}.

HiOp has been architected for parallel execution. The optimization algorithm in HiOp is implemented using a
generic programming approach where all linear algebra operations are called through an abstract interface
independent of underlying data structures and details of linear algebra kernels implementations.
This allows different parallel implementations of linear algebra kernels to be added to HiOp without modifying the
optimization algorithm code. Kernels K1, K2 and K3 are implemented as methods of vector, dense matrix and sparse matrix
classes.

\subsection{Umpire and RAJA Portability Libraries}

Hardware abstraction libraries, such as Umpire \cite{beckingsale2019umpire}, RAJA \cite{beckingsale2019raja}
or Kokkos \cite{CarterEdwards20143202} allow one to write code for heterogeneous hardware platforms independent of the vendor-specific details.
The hardware backend is selected simply by choosing appropriate template parameters.
The rest of the code remains unmodified, irrespective
of the device on which the code is run. We chose a programming model based on the RAJA and Umpire libraries because of portability,
relatively low training and development cost, code readability, and flexibility they provide when debugging and 
optimizing the code\footnote{We would like to note here that Kokkos provides very similar capability and the decision which one to pick was a difficult one.}. We discuss these in detail here, using simple examples.
Consider a dot product vector kernel:
\begin{lstlisting}[language=C++]
double dotProd(const double* x,
               const double* y, int N)
{
  double dot = 0;
  for(int i=0; i<N; ++i) {
    dot += x[i]*y[i];
  }
  return dot;
}
\end{lstlisting}
A RAJA implementation of the same kernel may look something like this:
\begin{lstlisting}[language=C++]
#define RAJA_LAMBDA [=]
using namespace RAJA;
using reduce_policy = RAJA::seq_reduce;
using exec_policy   = RAJA::seq_exec;
double dotProd(const double* x,
               const double* y, int N)
{
  ReduceSum<reduce_policy, double> dot(0.0);
  forall<exec_policy>(RangeSegment(0, N),
    RAJA_LAMBDA (Index_type i) {
      dot += x[i]*y[i];
    });
  return dot.get();
}
\end{lstlisting}
While the setup, predicate and next condition of the loop look a bit different, the body of the loop remains unchanged. Once the
loop body is verified for correctness, it can be built and executed on different hardware simply by changing
policies \verb|seq_reduce| and \verb|seq_exec|. Of course, the loop body must be implemented in
a thread-safe fashion in the first place.

The RAJA code is self-documenting. An object of type \verb|ReduceSum| is where the sum reduction will
take place. The first template parameter specifies the reduction policy (sequential) and
the second one specifies the data type of the summation result (double). A function template \verb|forall|
implements the for-loop. The template parameter \verb|seq_exec| specifies how the for-loop is
parallelized, the \verb|RangeSegment| object specifies the loop iteration pattern and the lambda function
passed as the second parameter defines the body of the loop.

Minimal code modification is needed to re-target the dot product function to operate on an NVIDIA GPU:
\begin{lstlisting}[language=C++]
#define RAJA_LAMBDA [=] __device__
using reduce_policy = RAJA::cuda_reduce;
using exec_policy   = RAJA::cuda_exec<256>;
\end{lstlisting}
We simply replace sequential policies with CUDA equivalents. The integer template parameter of the CUDA policy
determines CUDA thread block size. Another notable modification is that
the lambda function needs \verb|__device__| decoration to be correctly compiled by CUDA.

The RAJA library hides a number of implementation details from the user. For example, a
CUDA implementation of the dot product may use shared memory or move last few steps of the reduction back to the host (see reduction examples in \cite{nvidia2018sdk}). 
All these implementation details are hidden from the RAJA user. However, the user can select one of the CUDA policies available in RAJA library that is most suiteable for the problem at hand.

This software design allows for separation of responsibilities -- the applied math experts can work on
different kernel implementations while computer science experts can select appropriate execution
policies or contribute new policies to the RAJA library.

The RAJA lambda functions typically operate on raw data arrays. It is the responsibility of the RAJA user to ensure the
locations of the data are accessible from the selected RAJA execution policies. The
Umpire library \cite{beckingsale2019umpire} is a hardware abstraction layer for memory management
on machines with multiple memory devices. Among other capabilities, Umpire provides a portable
replacement for standard library functions such as \verb|malloc|, \verb|free| and
\verb|memcpy|. The central concept in Umpire is the resource manager, which is implemented as a singleton
that can be accessed through a local reference in the code. The resource manager can be used to
create allocators or copy data between memory spaces. 
\begin{lstlisting}[language=C++]
auto& rm = umpire::ResourceManager::getInstance();
umpire::Allocator al = rm.getAllocator("HOST");
double* data = static_cast<double*>(al.allocate(N*sizeof(double)));
// some code ...
al.deallocate(data);
\end{lstlisting}
In this example, the resource manager was accessed by reference in a local scope and used
to create an allocator for host memory space. The allocator is then used to allocate an array of
$N$ double precision elements. 
much larger. 
The \verb|deallocate| method of the allocator class corresponds to the \verb|free| call from the standard library. When Umpire is built with CUDA backend, allocators for
pinned, device, and unified virtual memory can be created, as well.

The resource manager also provides methods to copy and move Umpire arrays. These methods can copy/move
data within the same or between different memory spaces for as long as the data is registered with the
resource manager.

\subsection{Refactoring HiOp Code}

When porting HiOp code to hardware accelerators, we were guided by following principles: ($i$) make minimal
changes to HiOp code to reduce possibility of introducing new bugs and ($ii$) do not replace existing classes,
but implement additional RAJA-based linear algebra classes. Keeping existing HiOp linear algebra intact provides
us a mature and well-tested reference against which to verify the RAJA-based implementation.
It also makes RAJA and Umpire optional dependencies, so that the prior usability of HiOp is not
diminished by new dependency requirements.

The linear algebra module in HiOp already offered base classes for vectors and matrices, so we were 
able to add RAJA-based equivalents to existing linear algebra implementations seamlessly.
All RAJA-based classes were implemented to use Umpire for memory management.
We added string parameters to constructors of RAJA-based classes to pass memory space
specifier for Umpire. If the selected memory space is \verb|DEVICE|, the constructor will
create a host mirror in addition to allocating memory on an accelerator device. We also added
methods to linear algebra base classes to copy data between the device and the host mirror.

While Umpire allows for a runtime decision about memory space, RAJA execution policies are
specified at compile time. We included a stanza like the following code fragment at the beginning of each source file
containing the definition of RAJA-based classes:
\begin{lstlisting}[language=C++]
#ifdef HIOP_USE_GPU
  #include "cuda.h"
  #define RAJA_LAMBDA [=] __device__
  using hiop_raja_exec   = RAJA::cuda_exec<128>;
  using hiop_raja_reduce = RAJA::cuda_reduce;
  using hiop_raja_atomic = RAJA::cuda_atomic;
#else
  #define RAJA_LAMBDA [=]
  using hiop_raja_exec   = RAJA::omp_parallel_for_exec;
  using hiop_raja_reduce = RAJA::omp_reduce;
  using hiop_raja_atomic = RAJA::omp_atomic;
#endif
\end{lstlisting}
Note that we need to use a preprocessor directive to add \verb|__device__| decoration to
the lambda-function when GPU support is enabled. This is a less flexible solution
than setting aliases to policy types with \verb|using| keyword.

\subsubsection{Linear algebra factory}

Some minor refactoring was required to eliminate implementation-specific code from
HiOp's optimization algorithm classes. We implemented a linear algebra factory class to
construct polymorphic linear algebra objects and replaced direct calls to constructors.
The factory is implemented as a static class with the member variable \verb|mem_space_|,
which is set by a user-selected option. In addition to four memory space options used
by Umpire, we added another \verb|DEFAULT| option, which we use to select HiOp's legacy
linear algebra objects. If HiOp is built without RAJA support, only the default
option is available to the user. 

Based on user selection, either a legacy or a RAJA-based vector is created and a
handle to it is returned as a pointer to the base vector class \verb|hiopVector|.
By this relatively simple modification, we ensured all linear algebra objects in
HiOp are accessed through their abstract interfaces.

\subsubsection{Dense matrix implementation}

HiOp's dense matrix class stores matrix data in a contiguous memory block in a row major format. The sequential dense matrix implementation provides an array of row pointers of type \verb|double**| to allow for intuitive access to matrix elements (e.g. \verb|data[row][col]|). In our RAJA implementation of the dense matrix class, we use \verb|RAJA::View| to access matrix elements, which is a more flexible and configurable approach.

To illustrate how using a \verb|RAJA::View| simplifies implementation, consider a dense matrix-matrix multiplication kernel $Y \leftarrow \beta Y + \alpha A X$, where $Y$ is $m \times n$, $M$ is $m \times q$ and $X$ is $q \times n$ matrix. Assuming \verb|ydat|, \verb|adat| and \verb|xdat| are pointers to the data of matrix instances $Y$, $A$ and $X$, respectively, such kernel can be implemented in RAJA like this:
\begin{lstlisting}[language=C++]
using namespace RAJA;
View<double, Layout<2>> Yview(ydat, m, n);
View<double, Layout<2>> Aview(adat, m, q);
View<double, Layout<2>> Xview(xdat, q, n);
RangeSegment row_rng(0, m), col_rng(0, n);
kernel<matrix_exec>(make_tuple(col_rng, row_rng),
  RAJA_LAMBDA(int col, int row)
  {
    double sum = 0;
    for (int k = 0; k < q; k++)
      sum += Aview(row, k) * Xview(k, col);
    Yview(row, col) = beta*Yview(row, col)
                      + alpha*sum;
  });
\end{lstlisting}

When constructing a \verb|RAJA::View|, the template parameters define the pointer type and the layout type, with the constructor parameters being the pointer to the data itself, followed by the size of each dimension of the matrix. Here, the layout is only defining the number of dimensions, with the right most index indicating entries in contiguous memory. Since the dense matrix class in HiOp stores data in a row major format, this default \texttt{RAJA::View} is sufficient for our application. Once the View is created, the underlying data can be accessed using the parentheses operator to enable intuitive matrix element accessing.

The \verb|matrix_exec| used in the preceding code snippet is a custom RAJA kernel policy that we defined in addition to the aforementioned RAJA excution policies at the top of the dense matrix header file:

\begin{lstlisting}[language=C++]
using namespace RAJA;
#ifdef HIOP_USE_GPU
// Other RAJA execution policies ...
using matrix_exec =
KernelPolicy<
  statement::CudaKernel<
    statement::For<1, cuda_block_x_loop,
      statement::For<0, cuda_thread_x_loop,
        statement::Lambda<0>
      >
    >
  >
>;
#else
// ...
using matrix_exec = 
KernelPolicy<
  statement::For<1, hiop_raja_exec,   // row
    statement::For<0, hiop_raja_exec, // col
      statement::Lambda<0> 
    >
  >
>;
#endif
\end{lstlisting}
A kernel policy is a list of \verb|RAJA::statement|s, so here we define a matrix execution policy comprised of one \verb|RAJA::statement::For| that is linked to index 1 of the iteration space (the \verb|row_rng RangeSegment| from above), which calls another \verb|RAJA::statement::For| that is linked to index 2 of the iteration space (the col\_rng RangeStatement), which in turn makes a call to the main \verb|RAJA::statement:Lambda|. The execution policy for each \verb|RAJA::statement::For| can also be customized per the execution space, enabling the same flexibilty as with other RAJA execution policies.
Other storage formats (column major, blocked matrix) are configurable with different types of views. For example, to modify the previous code snippets to view data stored in a column major format, one only needs to flip 1 and 0 template parameters from the two \texttt{statement::For} templates. Interested reader can find additional examples in \cite{beckingsale2019raja} and references therein.

Not only does it simplify the implementation of complex linear algebra kernels, but it also produces readable and self-documenting code that is easy to understand and debug. Trying to implement more complex kernels that involve multiple input matrices, such as multiplying the transpose of one matrix by a constant, and then adding it to upper triangle of another matrix, was made trivial through tools such as the \verb|RAJA::View|. RAJA can also be built with the CMake variable \verb|RAJA_ENABLE_BOUNDS_CHECK| to get run-time bounds checking on RAJA Views.

\subsubsection{Mixed dense-sparse matrix implementation}

Mixed dense-sparse linear algebra is used to create a compressed linear system that
can be efficiently solved using dense linear solvers. A dense-sparse matrix
object \verb|hiopMatrixMDS|, which inherits directly from HiOp's base matrix class,
and its symmetric counterpart \verb|hiopMatrixSymBlockDiagMDS|, are central to that concept.

The mixed dense sparse matrix is essentially a container for a dense and a sparse
Jacobian matrices $J_{x_d}g$ and $J_{x_s}g$ in blocks $(3,1)$ and $(3,2)$, as well
as Jacobian matrices $J_{x_d}h$ and $J_{x_s}h$ in blocks $(4,1)$ and $(4,2)$ of the KKT
matrix in (\ref{linsys4x4}), respectively. The symmetric (block diagonal) dense-sparse
matrix is used as a container for matrices $Q_{x_s}$ and $Q_{x_d}$ in blocks $(1,1)$ and
$(2,2)$ of the KKT matrix. Current implementation of mixed dense-sparse linear algebra does not
support distributed memory parallelism.

In the original HiOp implementation, mixed dense-sparse matrices contained pointers to
specific implementations of dense and sparse matrices. Rather than creating a RAJA version
of mixed dense-sparse matrix objects, we refactored matrix interfaces to create abstract
dense and sparse matrix classes. With a minor amount of refactoring, we made mixed dense-sparse matrix
classes agnostic of implementations of their dense and sparse matrix members. The only
significant modification was the replacement of constructors for dense and sparse matrices with
factories.
\begin{lstlisting}[language=C++]
class hiopMatrixMDS : public hiopMatrix
{
private:
  hiopMatrixSparse* mSp;
  hiopMatrixDense*  mDe;
  // more member variables ...
public:
  // implemented public methods ...
  
  void timesVec(
    double beta,  hiopVector& y,
    double alpha, const hiopVector& x) const
  {
    assert(x.get_size() == mSp->n()+mDe->n());
    double* yd = y.local_data();
    const double* xd = x.local_data_const();
    mSp->timesVec(beta, yd, alpha, xd);
    mDe->timesVec(1.,   yd, alpha, xd+mSp->n());
  }
};
\end{lstlisting}
HiOp does not have a subvector object, so some of the mixed dense-sparse operations
access raw vector data and use pointer arithmetic to separate sparse and dense vector
elements, as shown in
the code snippet above. This violates data encapsulation for the vector, but is
GPU portable for as long as vector data is allocated in a contiguous memory block. The method \verb|local_data|
returns a pointer to vector local data in the memory space selected in the linear
algebra factory. Since the factory is static and the memory space is set at the setup
stage through user options, all linear objects will use the same memory space.

Compression of (\ref{linsys4x4}) to (\ref{eq:compressed}) is performed using fused kernels K3  of type $M := M+A Q_{x_s}^{-1} B^T$ (see Sec.~\ref{sec:complinalg}). Here $M$ is the dense matrix (\ref{eq:compressed}), $A$ and $B$ are sparse Jacobians, and $Q_{x_s}$ is stored as a vector since we assume it is diagonal. That way (2,2), (2,3) and (3,3) blocks in (\ref{eq:compressed}) are computed on GPU with one kernel launch each. Performance of these kernels could be improved if sparse matrices are stored in compressed sparse row instead of triplet format that HiOp uses, however that would require a significant rewrite of sparse linear algebra in HiOp.


\subsection{Porting Lessons Learned} \label{sec:lessons}

Incremental verifiability was crucial for porting HiOp to GPU -- we heavily depended on our unit testing framework when porting the linear algebra kernels. Unit testing allowed us to port and verify linear algebra kernels one-by-one without invoking optimization algorithm classes, and also to ``parallelize'' the development effort with multiple developers working on different kernels simultaneously. The three most time-consuming bugs to fix were in the cases where preconditions for linear algebra kernels were incorrectly documented and, as a result, unit tests were not correctly designed giving false positives.

HiOp is designed so it encapsulates all of its parallelizable code within linear algebra kernels, so porting those to GPU should in principle deliver fully GPU-enabled HiOp implementation. In practice, moving from testing standalone kernels to running proxy apps that test the entire algorithm is when most of bugs show up. Using a programming model based on Umpire and RAJA enables a three-stage approach to debugging: run the optimization engine (a) entirely with host execution policies and data residing on the host, (b) with device execution policies and data in unified virtual memory, and (c) with device execution policies and the data on the device. All three stages were executed with the same code, only replacing memory space parameters at runtime and execution policies at compile time.

We were able to catch and fix most of the bugs at the stage (a), using OpenMP execution policies. At that stage we were able to use host debuggers, which are more mature and feature rich than those for GPUs, and that sped up debugging significantly. At  stage (b), we were left with CUDA specific bugs only. Most notably, lambda functions used in RAJA kernels cannot capture class member variables when using CUDA execution policies and throw CUDA memory access exceptions. Instead, one needs to create a local copy in the method calling the RAJA kernel. A simple illustration of this bug is shown in a code snippet from \verb|hiopVectorRajaPar| below:
\begin{lstlisting}[language=C++]
using namespace RAJA;
// Many implementation details omitted
class hiopVectorRajaPar {
  double* data_;
  int N_;
  void addConstant(double);
};

void hiopVectorRajaPar::addConstant(double c) {
  double* data = data_;
  forall<cuda_exec<256>>(RangeSegment(0, N_),
    [=](Index_type i) {
      // data_[i] += c; // <- throws exception
      data[i] += c;
    });
}
\end{lstlisting}
In our experience, this issue seems to appear indeterministically -- the compiler sometimes can produce a well functioning binary. Because of that, it is difficult to create unit tests that catches this issue.  

Using unified virtual memory at the second debugging stage helped us avoid dealing with bugs that are more time consuming to fix, such as those where the optimization algorithm tries to access raw data directly on the host. These are instances where data was not correctly encapsulated in HiOp implementation. They went undetected because they did not introduce bugs or impact performance of {\em sequential} computations. Fixing this requires abstracting these computations out of the optimization algorithm and creating new specialized linear algebra kernels. One such example we found in \verb|addLinearDampingTermToGrad_x| method in \verb|hiopIterate| class.

There were a few smaller issues with our expectations of the portability libraries as well. For example, \verb|memcpy| from standard library does not check the size of the source and destination arrays before copying. Some edge cases in HiOp code would ``copy'' zero elements from a zero length array. On the other hand, Umpire's equivalent of \verb|memcpy| throws an exception if the Umpire array is of zero length, so we needed to modify code to handle those edge cases separately.

We also encountered issues with the legacy interface, which has been designed with an assumption that all data resides on the host. Dense matrix data is passed to model interface as an array of row pointers of type \verb|double** data|, which allows for intuitive indexing of matrix elements by the model callback function (e.g. \verb|data[row][col]|). For backwards compatibility, we created an array of row pointers in the Umpire memory space as a member of RAJA dense matrix class. This solution is not quite portable. For example, if Umpire memory space is on the device, one cannot get pointer to matrix data on the host by dereferencing \verb|data[0]|. A better solution would be to modify HiOp model interface to pass dense matrix data by a pointer \verb|double* data| to the model callback function. That would also give users more flexibility to create custom matrix views at the model side.

\section{Performance Evaluation} \label{sec:perfeval}



\subsection{Hardware Setup}
The performance evaluation was performed using an NVIDIA Tesla V100 GPU with 16 GB RAM, an Intel(R) Xeon(R) Silver 4110 CPU, with 256GB of RAM, running a Linux operating system.  The GPU is of exactly the same type as used in one of our targeted supercomputing systems, which is the Summit supercomputer at the Oak Ridge Leadership Computing Facility.  Although we have access to the supercomputer for other experiments, we used the specially built experimental machine with the same GPU as that of Summit in order to be able to execute profiled executions that consume a very long time due to profiling overheads.

\subsection{Software Setup: Dense Linear Algebra}
To measure the quantitative gains in performance obtained from porting to accelerator-based hardware architectures, we used three different dense linear algebra implementations.  Specifically, we used the following implementations of the Basic Linear Algebra Sub-programs (BLAS):
\begin{enumerate}
    \item OpenBLAS: This is used as an open-source, CPU-based baseline implementation commonly available for multiple CPU types and operating systems.
    \item Intel MKL: This is used as one of the fastest vendor-optimized, CPU-based dense linear algebra solvers, specifically designed by the vendor for Intel platforms.
    \item MAGMA: This is the accelerator-based dense linear algebra solver we used for GPU-based execution, which in turn uses CUBLAS underneath for BLAS operations on NVIDIA GPUs.
\end{enumerate}

\subsection{Software Setup: Timing}
In order to gather detailed timing information to measure the run time performance of our solver, we analyzed the code and manually introduced wall-clock timers across different parts of the functions to measure elapsed time for each function for each \HiOp iteration. As every iteration of \HiOp solver is nearly similar in behavior to other iterations, we present the average timing information per iteration for each experiment conducted.

For controlled experiments in which matrix sizes can be systematically varied, we used a \HiOp mini app, \texttt{nlpMDS\_ex4}, to measure the CPU-based and GPU-based performance. The mini app solves a convex optimization problem with $m=n_s+3$ constraints. In our tests, we set $n_d=n_s=k$ and varied $k$ from $2,000$ to $22,000$. The compressed matrix (\ref{eq:compressed}) of the model implemented in the mini app has size $(2k + 3) \times (2k + 3)$.

We analyzed the code and placed manually inserted wall-clock timers across the core solver parts that perform the matrix multiplication operations. Before collecting the wall-clock timings, a GPU synchronization API is called to ensure all pending GPU kernels are completed. For more detailed analysis of routines beyond the primary linear algebra solver, we also employed multiple profiling tools including \HPCToolkit and NVIDIA \texttt{nvprof}/\texttt{nvvp}.  The coarse units are more readily profiled using our manual timing instrumentation while the finer timing details are obtained using the automatic profiling tools.  We have found that the manual timing instrumentation is the fastest way to obtain the timing, whereas the automatic profiling tools introduced very large amount of profiling time overheads, often taking $100\times$ or even $1000\times$ more time to complete than a normal unprofiled run.

\subsection{Summary of Linear Algebra Performance}
From the runtime experiments, it is observed that GPU-based MAGMA execution delivers significant reduction in run times, as compared to both OpenBLAS as well as Intel MKL. On a matrix of size 16,000, OpenBLAS takes 72.95 seconds, Intel MKL takes 18.56 seconds, and MAGMA takes 4.49 seconds per iteration. In essence, GPU-based MAGMA execution is observed to offer $16\times$ and $4\times$ faster execution than OpenBLAS-based and Intel MKL-based execution, respectively. As the predominant consumer of computational time, the core of the solver depends on the factorization of a symmetric matrix, namely, the function \texttt{DSYTRF}, which computes the factorization of a real symmetric matrix $A$ using the Bunch-Kaufman diagonal pivoting method.

\begin{table}[htbp!]
\centering
\begin{tabular}{|l|r|r|}\hline
\textbf{Solver} & \textbf{Time} & \textbf{Percentage}\\\hline\hline
OpenBLAS & 67.92s & 93\% \\\hline
Intel MKL & 13.61s & 73\% \\\hline
MAGMA & 0.81s & 18\% \\\hline
\end{tabular}
\caption{Time taken by dense linear solver (K4). First column shows average linear solver wall time per HiOp iteration; the second column shows fraction of the total HiOp wall time taken the by linear solver.}
\label{tab:dense-timings-comparison}
\end{table}

The time taken per iteration of the solver with each linear algebra implementation is shown in Table~\ref{tab:dense-timings-comparison}. Therefore, clear gain in the speed of factorization is achieved using the GPU. The computation on the GPU basically consists of two parts: (1) copying data back and forth between  CPU and GPU, and  (2) actual computation of GPU factorization. Of the 0.81 seconds, about 0.45 is spent on memory copying, and the remaining 0.36 seconds are used for actual computation. This supports our rationale for moving the entire computation to GPU using RAJA portability library.

\subsection{Summary of RAJA-based Performance Gains}

\begin{figure}[!htbp]
    \centering
    \includegraphics[width=0.9\columnwidth]{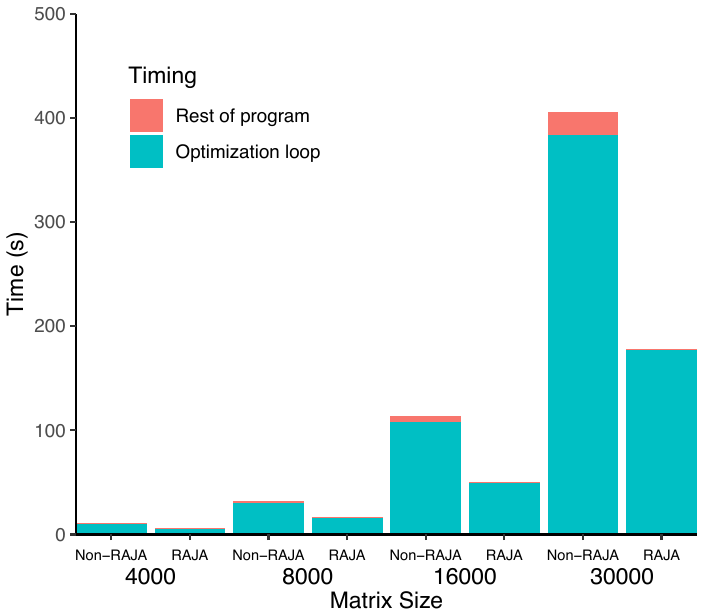}
    \caption{Performance before and after parallelization with RAJA}
    \label{fig:raja-vs-nonraja-runtime}
\end{figure}

Figure~\ref{fig:raja-vs-nonraja-runtime} shows the reduction in run times achieved by porting \HiOp memory and computation to GPUs with RAJA.  The matrix size is varied from small to the largest problem size that can effectively fit within the GPU memory limit of 16GB.  The timing labeled Non-RAJA corresponds to the best run time obtained by exploiting the GPU for dense linear algebra kernels only.  The timing labeled RAJA corresponds to Non-RAJA version further optimized with RAJA-based parallelization of most other operations.  The timing of the RAJA version is inclusive of all memory setup costs for transferring data between CPU and GPU memory in order for data to be consistent and synchronized across multiple RAJA kernels executed on the GPU.

As the matrix size increases, the overheads in non-RAJA were over $5\%$, while that in RAJA was significantly reduced to less than $0.5\%$ of total time.  Although RAJA-based execution involves new overhead of additional transfer data movement between the host and device, this is significantly offset by improved kernel performance from execution on the GPU. Furthermore, we expect that the entire host-device data movement overhead will be removed in RAJA implementation once minor design flaws in HiOp are addressed. The overheads occur primarily because Bunch-Kaufman solver in Magma has only host interface even though the computation is performed on the device. When the entire computation is on the device, the system matrix and the right hand side need to be moved back to the host only for Magma to pick it up and move it to the device again. A device interface for Bunch-Kaufman solver is under development and will be added in a future release of Magma. 

\begin{table}[htbp!]
\centering
\begin{tabular}{|l|r|r|r|}\hline
Version  &   SetMatrix  &   GetMatrix  &   TotalTransfer\\\hline\hline
Non-RAJA  &   6.4s  &   3.7s & 10.1s \\\hline
RAJA  &   10.8s  &   7.9s  & 18.7s \\\hline
\end{tabular}
\caption{Data movement time costs with RAJA vs. Non-RAJA versions. We expect most, if not all, host-device data movement to be eliminated in RAJA version after issues described in Sec.~\ref{sec:lessons} are addressed.}
\label{tab:transfertimes}
\end{table}

\remove{ 
\subsection{Detailed Hierarchical Timing Information}
Figure~\ref{fig:hiop-open-blas} shows the hierarchical call lists of major function per iteration. The most significant functions are highlighted in bold.
Figure~\ref{fig:hiop-mkl} shows the corresponding information for Intel MKL BLAS.  Figure~\ref{fig:hiop-magma} shows the information for MAGMA.

 \begin{figure}[!htbp]
    \centering
    \includegraphics[width=\columnwidth]{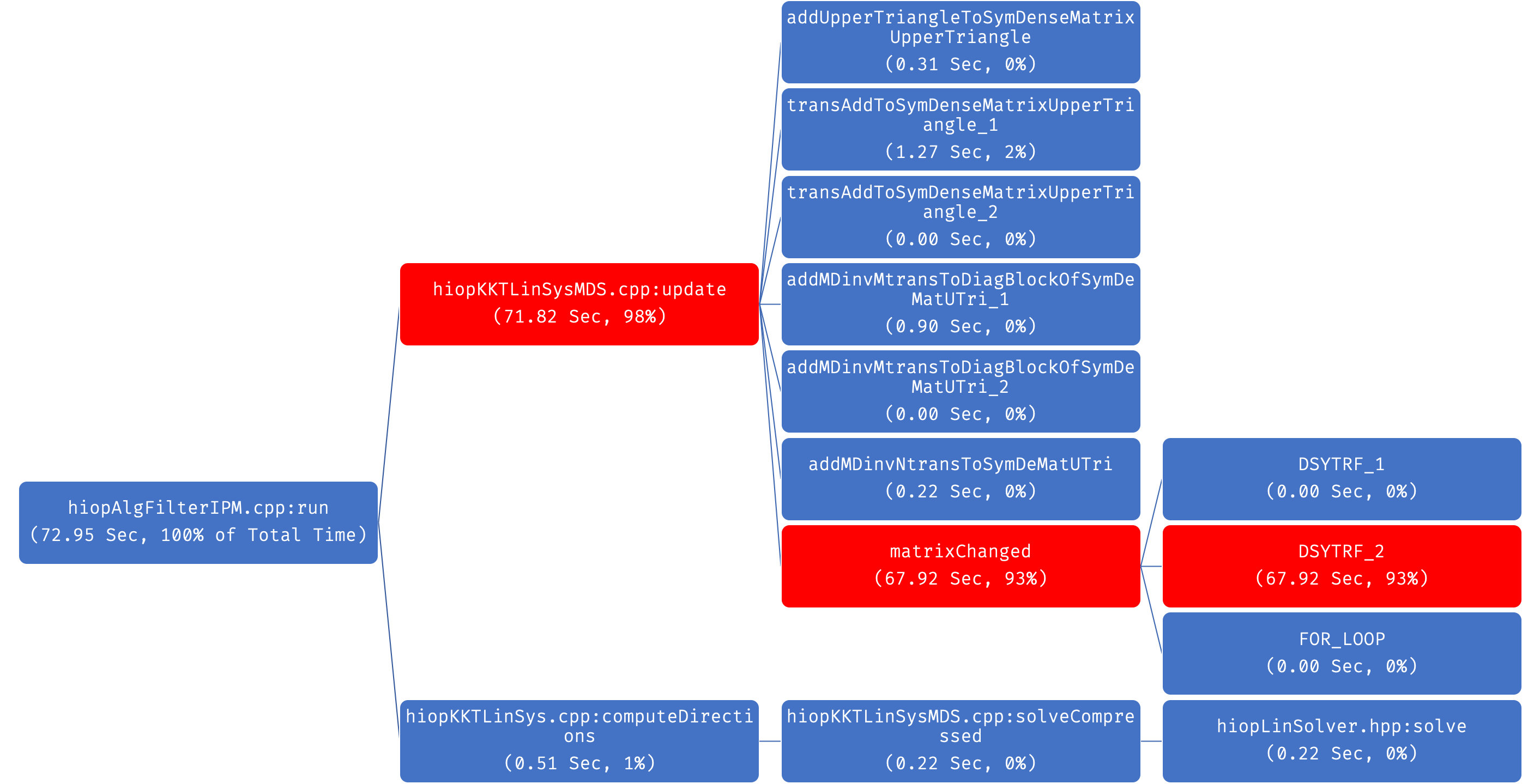}
    \caption{\HiOp Exploration}
    \label{fig:hiop-open-blas}
\end{figure}

 \begin{figure}[!htbp]
    \centering
    \includegraphics[width=\columnwidth]{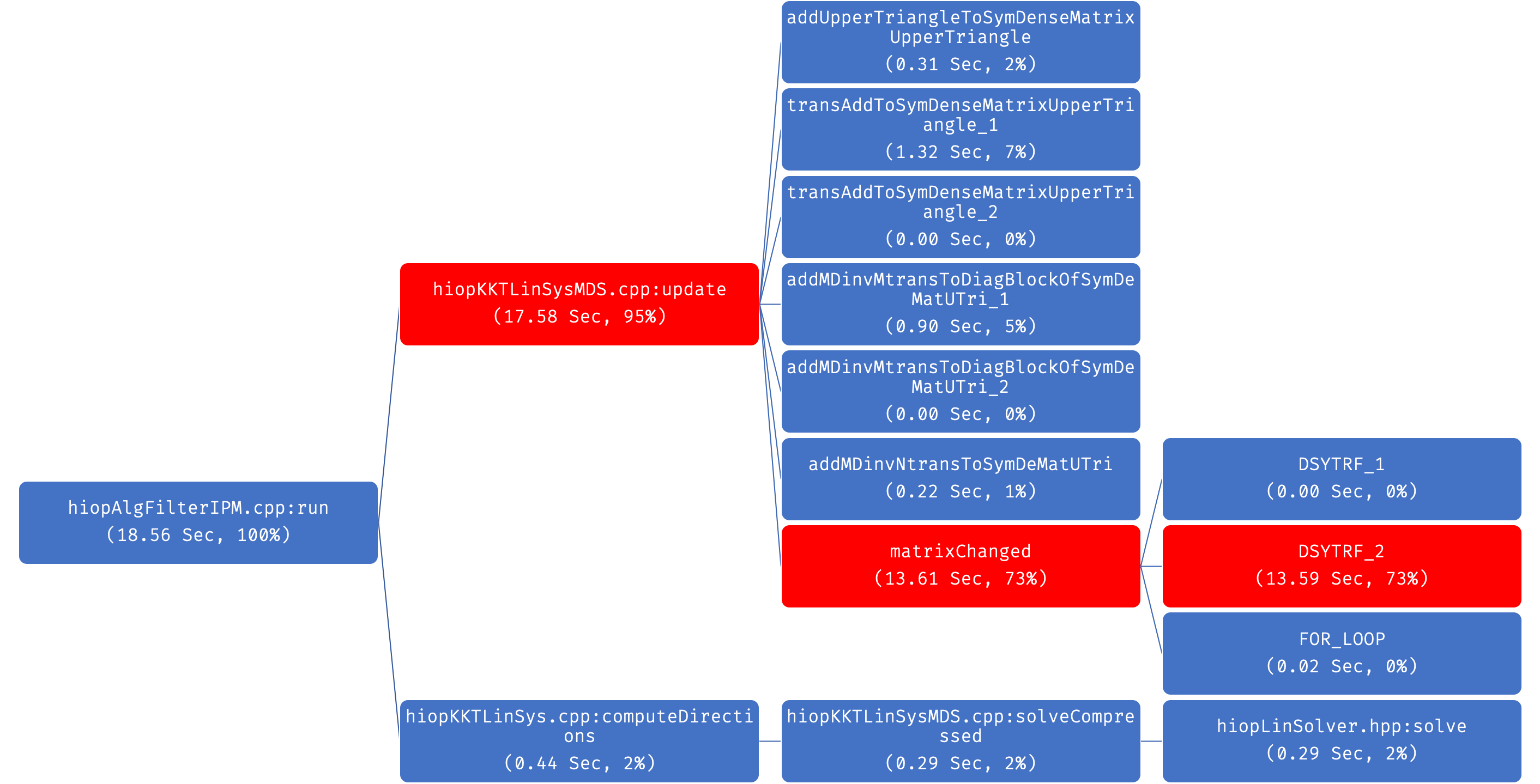}
    \caption{\HiOp Exploration}
    \label{fig:hiop-mkl}
\end{figure}

\begin{figure*}[!htbp]
    \centering
    \includegraphics[width=2\columnwidth]{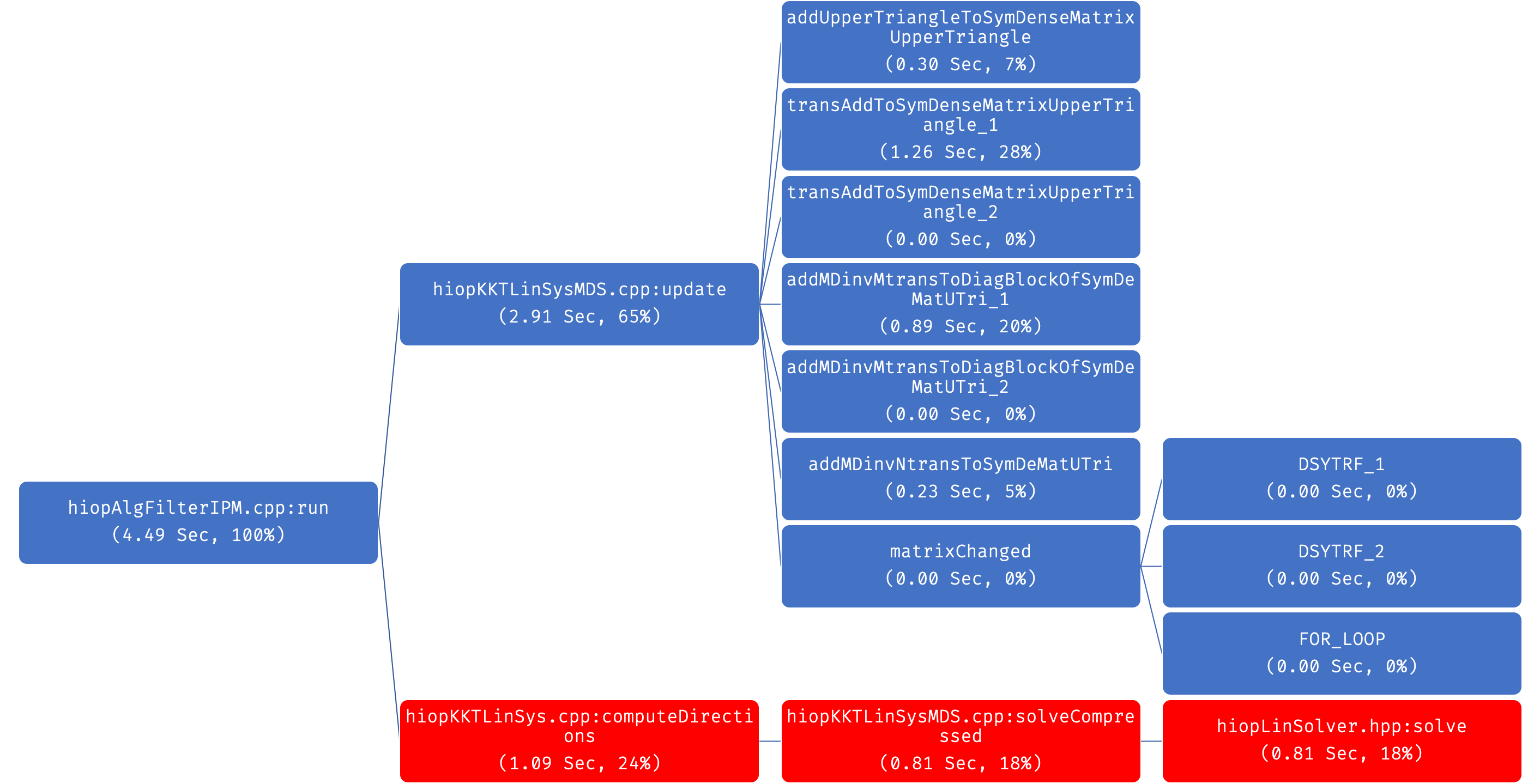}
    \caption{\HiOp Exploration}
    \label{fig:hiop-magma}
\end{figure*}
}

\subsection{Roofline Analysis}

We analyzed the performance of the \HiOp solver using a roofline analysis in which we plot the performance of the application based on the arithmetic intensity and device capability. 

The base roofline is obtained by the Empirical Roofline Toolkit (ERT), which runs a variety of micro-kernels and sweeps through a range of runtime configurations.  We compute the arithmetic intensity $\text{AI}$ as:

$$\text{AI} = \tfrac{\text{flop\_count\_dp}}{((\text{dram\_read\_transactions} + \text{dram\_write\_transactions})\times 32)}$$

where \texttt{flop\_count\_dp} is the total FLOP count for FP64 operations, and \texttt{dram\_read\_transactions} and \texttt{dram\_write\_transactions} are the read and write transactions from and to DRAM. The size of each memory transaction is 32 bytes, so the total DRAM data movement can be calculated as $(\texttt{dram\_read\_transactions} + \texttt{dram\_write\_transactions}) \times 32$B.

\begin{figure}[!htbp]
    \centering
    \includegraphics[width=1.0\columnwidth]{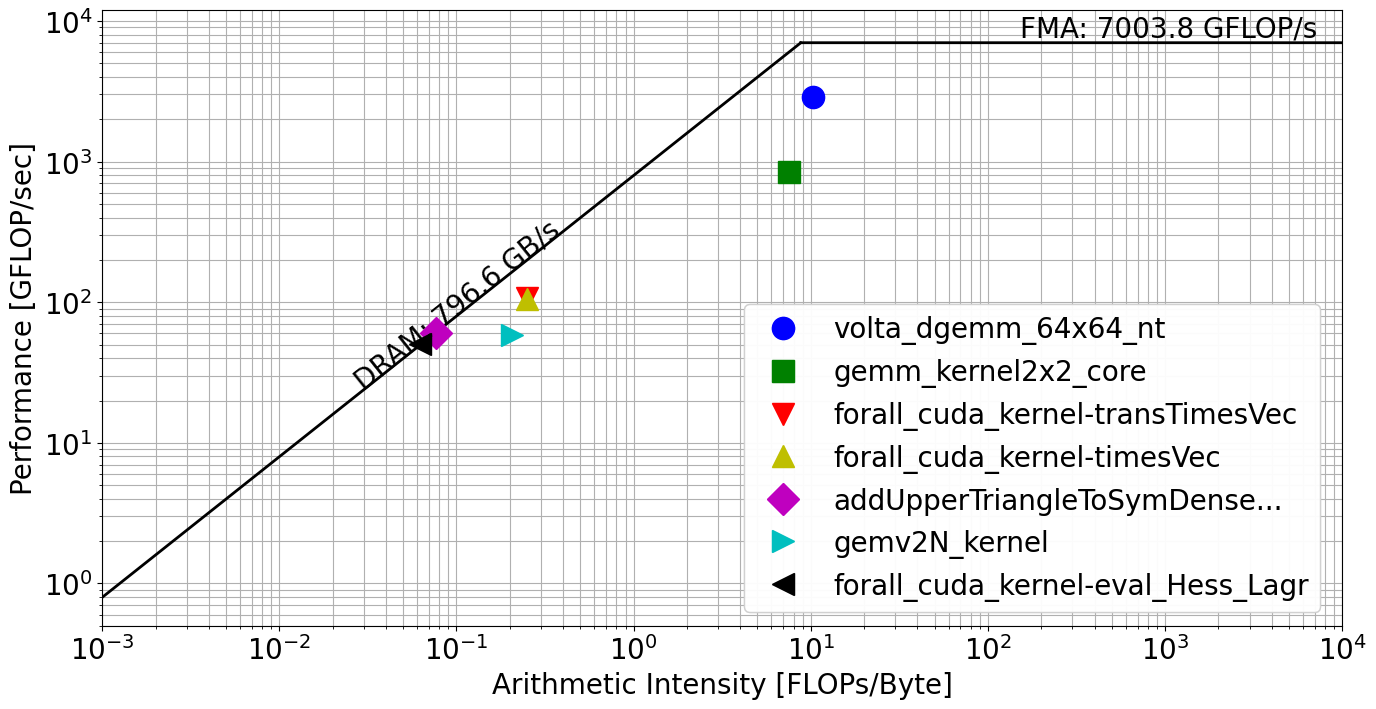}
    \caption{Roofline plot with RAJA-based execution}
    \label{fig:hiop-raja-roofline}
\end{figure}

The roofline plot for RAJA-based execution is shown in Figure~\ref{fig:hiop-raja-roofline}. The line in the plot represents the base roofline collected from running micro kernels on the device. The closer a point is to the line, the more optimized it is with respect to performance that can be achieved for the given arithmetic intensity.  Based on this observation, it can be seen that most of the routines are fairly optimized, with the possible exception of \texttt{gemm\_kernel2x2\_core} and \texttt{gemv2N\_kernel}.  Note that the sloped line labeled DRAM represents the limits of routines that are memory-bound.

Only the top few kernels of \HiOp are shown here, and others are omitted because they consume less than 0.5\% of the total execution time.  The performance of these individual computationally intensive kernels are plotted, including \texttt{volta\_dgemm\_64x64\_nt} and \texttt{gemm\_kernel2x2\_core} (matrix-matrix multiplication kernels called by MAGMA). As seen from the figure, the RAJA \texttt{forall} kernels seem to be memory bound and well optimized for their available computational intensity.  The dense linear algebra kernels also appear fairly optimized, tapping a high level of computational capability of the device.

For a matrix size of 20,000, the peak double precision floating point operation efficiency reached $60\%$, which corresponds to $4.2$ Tera FLOPs.  Almost all of this FLOP efficiency is achieved in the \texttt{volta\_dgemm\_64x64\_nt} routine.

\remove{ 
\begin{figure}[!htbp]
    \centering
    \includegraphics[width=\columnwidth]{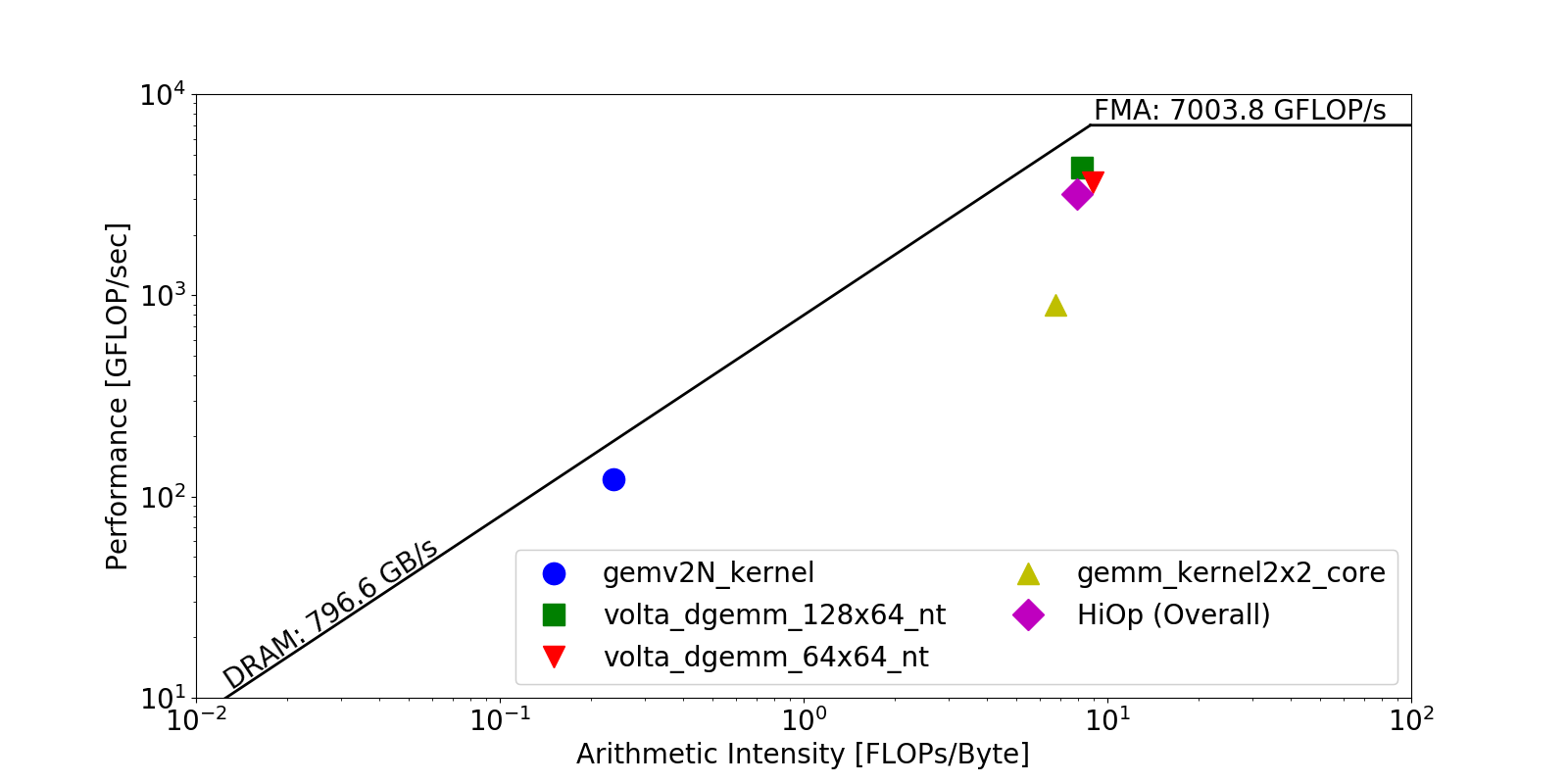}
    \caption{Roofline plot using \HiOp with a large matrix size of $30K\times 30K$}
    \label{fig:roofline-plot-15K}
\end{figure}

The roofline plot in Figure~\ref{fig:roofline-plot-15K} for a matrix size of 30,000 shows that \HiOp can perform exceedingly better with bigger matrix sizes and almost hitting the ceiling of the base roofline. Therefore, the kernels are observed to be well optimized.

\subsection{Time-Limited Partial Profiling}
Note that the profiling with smaller matrix size of $8K \times 8K$ took about $70$ hours for $8$ \HiOp iterations. Therefore, it could be safely assumed that profiling for larger matrix size will take a great amount of time to profile, which is not feasible most of the time. Furthermore, as the server gets busy it is not usable to run other GPU applications. For these reasons, we can perform time limited partial profiling to get some insight of the performance of the application. In this experiment, we used a larger matrix size of $30K \times 30K$ and executed the \nvprof profiler for $48$ hours to generate the roofline plot as shown in Figure~\ref{fig:roofline-plot-15K}.
\begin{figure}[!htbp]
    \centering
    \includegraphics[width=\columnwidth]{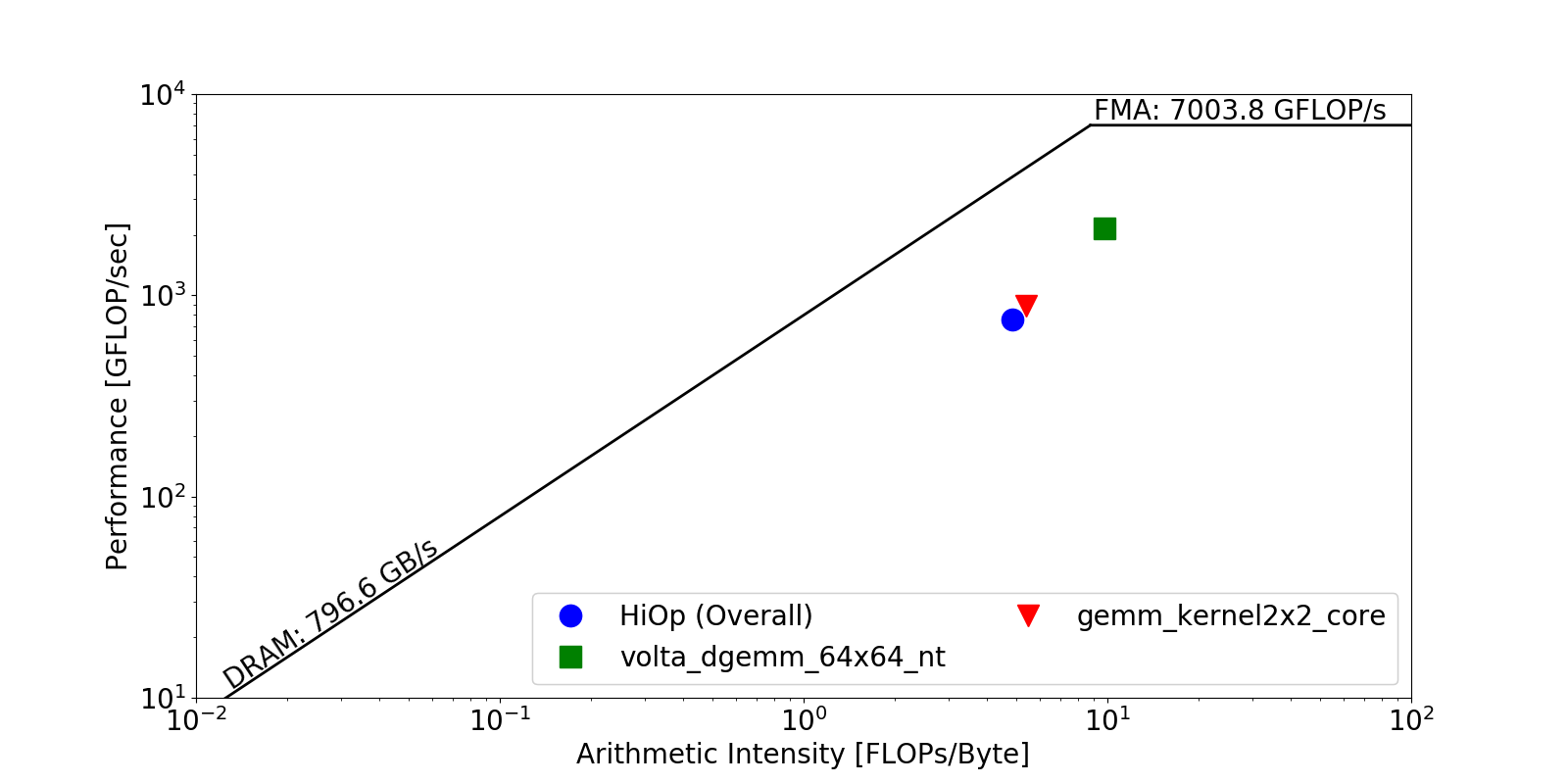}
    \caption{Roofline plot for matrix size of $8K\times 8K$}
    \label{fig:roofline-plot-4K}
\end{figure}
} 

\section{Conclusion} \label{sec:conclusion}

We have successfully ported HiOp optimization library to NVIDIA GPU hardware using Umpire and RAJA hardware abstraction libraries and leveraging Magma linear solvers. This effort included co-design of the optimization algorithm and the implementation in order to express the computation in terms of linear algebra kernels that are well suited for execution on hardware accelerators. The new algorithm compresses the sparse problem to a dense one, so we can use a dense linear solver from Magma library instead of a sparse one within the optimization workflow in Figure \ref{fig:workflow}. For problems of our interest, dense linear solvers show significantly higher hardware utilization on GPUs than presently available sparse linear solvers.

The porting to GPU was helped by HiOp's architecture, which provides data encapsulation and abstracts parallel code away from the optimization algorithm. Most of the porting work took place within HiOp's linear algebra module. In total 67 BLAS level-1, -2 and -3 kernels have been ported to RAJA and Umpire. Most of these kernels are specific to HiOp and tailored for the needs of the optimization algorithm. They fuse several simple linear algebra operations together for better performance. Porting this many kernels was greatly simplified by using the hardware abstraction libraries. In the process, we uncovered places in the code where HiOp departed from data encapsulation and linear algebra abstractions. We used simple workarounds there. Some minor code refactoring is required for a more sustainable solution and will be done in the next HiOp release.

The initial profiling results were quite encouraging as they showed significant speedup compared to the CPU-only solution, as well as CPU solution with the linear solver running on GPU. It is encouraging that the speedup improves as the size of the problem increases. Preliminary roofline analysis suggests moderate performance improvements are possible with optimizing certain HiOp kernels. Further improvements are possible with better computation orchestration and more aggressive problem compression. These findings suggest that mixed dense-sparse implementation of interior point method is a feasible approach for deploying nonlinear optimization on hardware accelerators. We believe this contribution is a solid baseline for future investigation of interior point method formulations suitable for execution on hardware accelerators.


\section*{Acknowledgements}
This research was supported by the Exascale Computing Project (17-SC-20-SC), a collaborative effort of the U.S. Department of Energy Office of Science and the National Nuclear Security Administration. 
We thank Chris Oehmen and Lori Ross O'Neil for critical reading of the manuscript and helpful suggestions. We also thank Tim Carlson and Kurt Glaesmann of Research Computing for their support (all from Pacific Northwest National Laboratory). Warm thanks go to Rich Hornung and David Beckingsale of Lawrence Livermore National Laboratory for their help with using RAJA and Umpire libraries. Finally, we acknowledge the support from the Oak Ridge Leadership Computing Facility, in particular a great deal of help from Philip Roth.

\bibliography{hiopbibfile}

\end{document}